\documentclass{llncs}
\usepackage{graphicx}
\usepackage{makeidx}
\begin{document}
\title{Decoherence in quantum spin systems}
\author{H. De Raedt$^1$ and V. V. Dobrovitski$^2$}
\institute{
$^1$Applied Physics - Computational Physics,\\
Materials Science Centre, University of Groningen\\
Nijenborgh 4, NL-9747 AG,Groningen, The Netherlands\\
E-mail: deraedt@phys.rug.nl; http://www.compphys.org\\
$^2$Ames Laboratory, Iowa State University, Ames IA 50011, USA\\
E-mail: slava@ameslab.gov\\
Version: \today
}
\maketitle
%
%
%
%%%%%%%%%%%%%%%%%%%%%%%%%%%%%%%%%%%%%%%%%%%%%%%%%%%%%%%%%%%%%%%%%%%%%%%%%%%%%
%
\begin{abstract}
Computer simulations
of decoherence in quantum spin systems require
the solution of the time-dependent Schr\"odinger equation 
for
interacting quantum spin 
systems
over extended periods of time.
We use exact diagonalization,
the Chebyshev polynomial technique,
four Suzuki-formula algorithms, and the short-iterative-Lanczos method
to solve a simple model for decoherence of a quantum
spin system by an environment consisting of quantum spins,
and compare advantages and limitations of
different algorithms.
\end{abstract}
%
%%%%%%%%%%%%%%%%%%%%%%%%%%%%%%%%%%%%%%%%%%%%%%%%%%%%%%%%%%%%%%%%%%%%%%%%%%%%%
%
%
%
%-------------------------------------------------------------
%
\section{Introduction}\label{sec1}
%
%-------------------------------------------------------------
%
The description of a quantum spin system (below referred to
as a central spin system (CSS)) interacting with its
quantum environment (bath) is among the most fundamental problems
of theoretical physics. Even if the energy exchange between
the CSS and the bath is absent (no dissipation), the system-bath
interaction still strongly affects the motion of the CSS
due to a loss of phase coherence between different eigenstates
of the central system. This many-body quantum phenomenon is 
commonly called decoherence. 

Decoherence is fundamental
for quantum measurement theory \cite{Blanchard,Zurek,Joos}
and for condensed matter physics; it
can suppress the tunneling of defects in crystals \cite{Legget},
spin tunneling in magnetic molecules and nanoparticles \cite{qtm94},
or can destroy
the Kondo effect in a dissipationless manner \cite{kondo}.
Decoherence is of particular relevance for quantum computation
since the loss of phase relations between different states of the
quantum computer may result in an accumulation of errors and
may prevent the computer from working correctly~\cite{Nielsen}.
A detailed theoretical understanding of decoherence
would definitely help to alleviate this fundamental problem.

Most theoretical studies of decoherence are based on a model
of a single spin interacting with a bath of bosons~\cite{Legget}.
This model is too simple in the context of e.g. quantum
computation or tunneling in magnetic molecules.
Extensive studies of many-spin central systems interacting
with other types of environment, such as a bath of nuclear
spins \cite{qtm94}, are needed.

A many-spin system interacting with a bath of quantum spins
presents a fairly complex many-body quantum problem,
and numerical simulation is an indispensable
tool for investigating the long-time 
dynamics of a decohered CSS.
One of the most reliable approaches is to model directly
the quantum motion of the whole system (CSS plus bath)
by solving the corresponding time-dependent Schr\"odinger
equation (TDSE).
For  such simulations, the numerical algorithms that solve
the TDSE should be 1) numerically stable (i.e. conserve the norm of the wave function)
for all integration times of interest,
2) sufficiently accurate and allow for controlled increase of 
the accuracy (e.g. to rule out that
the loss of phase coherence is due to poor accuracy, rounding errors etc.),
3) efficient in terms of memory and CPU use, in particular for large spin systems.
Below we compare three different numerical techniques that have the potential
to meet these requirements:
four Suzuki-formula algorithms~\cite{Suzuki77,DeRaedt87,Suzuki8591,hdr,Krech98,qce},
a Chebyshev polynomial technique~\cite{TAL-EZER,LEFOR,Iitaka01,Slava},
and the short-iterative-Lanczos method~\cite{LEFOR,Park,Manthe,Jaklic94}.
%
%-------------------------------------------------------------
%
\section{Model and algorithms}\label{sec2}

Consider the simple-looking but non-trivial model defined by the Hamiltonian
\begin{equation}
H=J_0({\vec S}_1+{\vec S}_2)^2 + \sum_{n=1}^{L} J_n \vec{I}_n\cdot({\vec S}_1+{\vec S}_2).
\label{ham}
\end{equation}
The CSS $({\vec S}_1,{\vec S}_2)$, where $S_1=S_2=1/2$, is coupled 
to $L$ bath spins $\{{\vec I}_n\}$ ($I_n=1/2$) by a Heisenberg exchange interactions $\{J_n$\}.
The initial states of the spins $\{{\vec I}_n\}$ are assumed to be random and
uncorrelated. For the initial state of the CSS we take
the state with one spin up and the other spin down.
We are interested in the time evolution of the magnetization of one of the CSS spins,
e.g. $\langle S_1^z(t)\rangle$.

A nice feature of the model (\ref{ham}) is that if
all $J_n=J$ then, in the large $L$ limit,
$\langle S_1^z(t)\rangle$ can be calculated exactly~\cite{Akakii}:
\begin{equation}
\langle S_1^z(t)\rangle=\frac{1}{6}[1+2(1-LJ^2t^2)e^{-LJ^2t^2/2}]\cos2(J_0-J)t.
\label{exact}
\end{equation}
The result (\ref{exact}) exhibits an interesting feature: initially,
the amplitude of the magnetization rapidly
decays to zero, then increases
again and becomes constant (1/6) as
$t\rightarrow\infty$~\cite{Akakii}.
This is similar to the two-step decoherence process discovered earlier~\cite{Slava01}
and can be understood from simple physical arguments
\cite{Akakii}.
The model (\ref{ham}) captures
some non-trivial aspects of decoherence,
and provides a simple test
to compare various algorithms for solving the TDSE
under conditions that are rather demanding from the point of view of algorithmic,
memory and CPU requirements.
We now discuss four different approaches to solve the TDSE for models such as (\ref{ham}).

{\bf Exact diagonalization} (ED) is the most straightforward approach. Standard library
routines can be used to compute all eigenvalues and eigenvectors of the $D\times D$ matrix $H$
($D=2^{L+2}$ denotes
the dimension of the Hilbert space spanned by the states of the $L+2$ spins
1/2).
The initial state is 
represented as a superposition of
eigenvectors, and the wave function $\psi(t)$ is obtained
by two matrix-vector multiplications of length $D$ and a phase-shift operation on a vector.
In practice, the amount of memory needed to store the $D\times D$ elements of the eigenvectors
limits the application of this approach to problems with $D$ of the order of 10000, which
corresponds to systems with about 14 $S=1/2$ spins.
Memory and CPU time of the 
ED algorithm scale as $D^2$ and $D^3$ respectively.

{\bf Suzuki product-formula algorithms} (SP) are based on the
approximation $e^{-i\tau H}\approx U_2(\tau)=e^{-i\tau H_1/2}\ldots e^{-i\tau H_p/2} e^{-i\tau H_p/2}\ldots e^{-i\tau H_1/2}$
where $H=\sum_{j=1}^p H_j$.
We consider two different decompositions that can be implemented efficiently: The original
pair-product split-up~\cite{Suzuki77,hdr} in which $H_j$ contains all contributions
of a particular pair of spins, and a XYZ decomposition in which we break
up the Hamiltonian according to the $x$,$y$ and $z$ components of the spin operators~\cite{qce}.
$U_2(\tau)$ is the building block for the fourth-order-in-time approximation
$e^{-i\tau H}\approx U_4(\tau)=U_2(a\tau)U_2(a\tau)U_2((1-4a)\tau)U_2(a\tau)U_2(a\tau)$
where $a=1/(4-4^{1/3})$~\cite{Suzuki8591}.
The error on the wave function is bounded as
$\Vert e^{-it H}\Psi(0)-U_n^m(\tau)\Psi(0)\Vert\le c_nt\tau^{n}$ where $t=m\tau$
and $c_n$ is positive constant.
By construction, all these algorithms conserve the norm of the wave function and, as
a consequence are unconditionally stable~\cite{DeRaedt87}.
These time-stepping algorithms advance the state of the quantum system
by small time steps $\tau$ ($\tau\Vert H\Vert\ll1$) and work equally well if the Hamiltonian
contains couplings to time-dependent external fields~\cite{qce}.
For a fixed accuracy,
memory and CPU time of the $n$-th order SP algorithm scales as $D$ and $nt^{(1+1/n)}D$ respectively

The {\bf Chebyshev polynomial algorithm} (CP)~\cite{TAL-EZER,LEFOR,Iitaka01,Slava} uses the identity
$\Psi(t)=\lim_{K\rightarrow\infty}\left[J_0(t\Vert H\Vert)I + 2\sum_{k=1}^{K} J_{k}(t\Vert H\Vert)\hat{T}_{k}(H/\Vert H\Vert)\right] \Psi(0)$.
The %matrix-valued modified Chebyshev
polynomials $\hat{T}_{k}(X)$ are defined by
the recursion $\hat{T}_{k+1}(X)\Psi(0)=-2iX\hat{T}_{k}(X)\Psi(0)+\hat{T}_{k-1}(X)\Psi(0)$ for $k\ge1$,
$\hat{T}_{0}(X)\Psi(0) = \Psi(0)$, and $\hat{T}_{1}(X)\Psi(0)=-iX\Psi(0)$.
Using standard 14-digit arithmetic, all Bessel functions $|J_{k}(z)|$
are zero to machine precision if $k>K=|z|+100=|t|\Vert H\Vert+100$ and therefore
the Chebyshev polynomial approximation to $\Psi(t)$ is accurate to machine precision
also (up to small rounding errors). 
Although the CP algorithm is not unconditionally stable,
it is so accurate that it can safely be used for time stepping (also with very large time steps).
Note that once $t$ has been fixed, the CP algorithm
cannot be used to generate reliable information for shorter times.
As $K$ is linear in $t$, the computation time required to reach a time $t$
increases {\sl linearly} with $t$ (and $D$).
This linear dependence on $t$ (and the very high accuracy) suggests that the Chebyshev polynomial algorithm
may be the method of choice if we want the solution of the TDSE for a few (very long) times~\cite{Slava}.
Memory and CPU time of the CP algorithm scale as $D$ and $tD$ respectively 
($K\ll D$ for most problems of interest).

The {\bf short iterative Lanczos algorithm} (SIL)~\cite{LEFOR,Park,Manthe,Jaklic94}
is based on the approximation
$e^{-i\tau H}\Psi\approx e^{-i\tau P_NHP_N}\Psi$
where $P_N$ is the projector on the $N$-dimensional subspace spanned by the vectors
$\{\Psi,H\Psi,\ldots,H^{N-1}\Psi\}$. We calculate $e^{-i\tau P_NHP_N}\Psi$ by generating
the orthogonal Lanczos vectors in the usual manner~\cite{Wilkinson}, and use exact diagonalization of the
resulting $N\times N$ tri-diagonal matrix 
for time propagation~\cite{LEFOR,Park,Manthe}.
Clearly $e^{-i\tau P_NHP_N}$ is unitary and hence the method is unconditionally stable.
The accuracy of this algorithm depends both on the order $N$ and the state 
$\Psi$~\cite{LEFOR,Park,Manthe}.
In exact arithmetic,
$e^{-i\tau H}=\lim_{N\rightarrow\infty} e^{-i\tau P_NHP_N}\Psi$, but in
practice, the loss of orthogonality during the Lanczos procedure~\cite{Wilkinson} limits the
order $N$ and the
time step $\tau$ that can be used without introducing spurious eigenvalues~\cite{Wilkinson}.
Furthermore, we require $N\ll D$ because the memory 
needed to store the eigenvectors
(and/or all Lanczos vectors) is proportional to $N^2$.
In practice, the low-order SIL algorithm may not work well if
$\Psi$ contains contributions from many
eigenstates of $H$ with very different energies,
because it is unlikely that all these eigenvalues will be
present in $P_NHP_N$ (for small $N$).
Memory and CPU time of the SIL algorithm scale as $D$ and $N^2Dt/\tau$ respectively.
In general, $N$ increases with $\tau$ in a non-trivial, problem dependent manner.

\section{Numerical tests}\label{sec3}

\begin{figure}[t]
\begin{center}
\setlength{\unitlength}{1cm}
\begin{picture}(6,4)
\put(-3.,0.){\includegraphics[width=7cm]{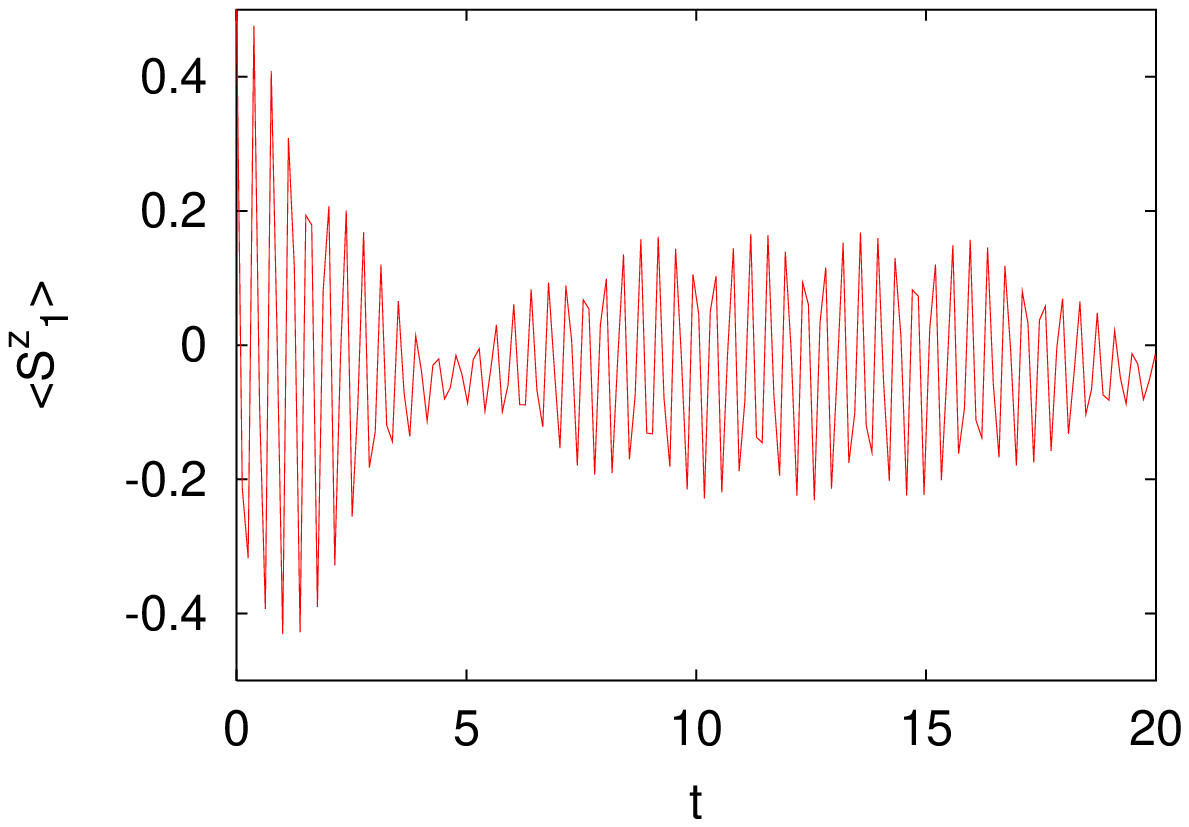}}
\put(4.,2.75){{\hsize=12truecm
\begin{tabular}{c|c|c}
Method& Error & CPU-time\\
\hline
\noalign{\vskip 2pt}
ED&-MP-&6739.1s\\
SP-Pair($U_2$)& $0.26\times10^{-3}$&$2.6$s\\
SP-Pair($U_4$)& $0.42\times10^{-8}$&$9.6$s\\
SP-XYZ($U_2$)& $0.97\times10^{-1}$&$1.1 $s\\
SP-XYZ($U_4$)& $0.23\times10^{-4}$&$5.6 $s\\
CP&-MP-&5.9s\\
SIL(N=5)& $0.29\times10^{-5}$&$68.3  $s\\
SIL(N=10)&-MP-&$137.8$s\\
\noalign{\vskip 2pt}
\hline
\end{tabular}
}}
\end{picture}
\caption{%
Left: Magnetization $\langle S_1^z(t)\rangle$ as a function of time as obtained by numerical
simulation of two central spins interacting with
a bath of $L=10$ spins. The parameters of model (\ref{ham}) are $J_0=8$, $J_k=0.128$.
Except for the CP algorithm, a time step $\tau=0.05$ was used.
Right:
Comparison of the efficiency of various algorithms
to solve the TDSE, for the case of the data shown at the left.
The entry -MP- denotes ``machine precision''.
CPU times as measured on a Windows 2000 Athlon XP 1900+ system.
}
\end{center}
\label{fig1}
\end{figure}

In Fig.~1, we show a typical simulation result for
$\langle S_1^z(t)\rangle$, as obtained by the CP solution of the TDSE for model (\ref{ham}).
The initial fast decay, and subsequent reappearance of the oscillations is clearly
present. Qualitatively these results agree with the analytical (large $L$) solution (\ref{exact}).
Also shown is the error $\Vert \Psi_{\hbox{ED}}(t=20) - \Psi_{\hbox{X}}(t=20)\Vert$ where X is
one of the seven algorithms used. It is clear that SIL is not competitive for this
type of TDSE problem, as already anticipated above. The fourth-order pair-approximation
is close but still less efficient than the CP algorithm, but the other SP algorithms are clearly
not competitive. The reason that the pair-approximation is performing fairly well in this
case is related to the form of the Hamiltonian (\ref{ham}).
The present results support our earlier finding~\cite{Slava} that
the numerical simulation of decoherence in spin systems is most efficiently done
in a two-step process: the CP algorithm can be used to make a big leap in time, followed
by the SP algorithm calculation to study the time dependence on a more detailed level.
From a more general perspective, to increase the confidence in numerical simulation results,
it is always good to have several different algorithms performing the same task.

%
%-------------------------------------------------------------
%
%\section*{Acknowledgements}
%
%-------------------------------------------------------------
%
\smallskip
This work is partially supported by the Dutch
`Stichting Nationale Computer Faciliteiten' (NCF).
This work was partially carried
out at the Ames Laboratory, which is operated for the U.\ S.\ Department of
Energy by Iowa State University under Contract No.\ W-7405-82 and was
supported by the Director of the Office of Science, Office of Basic Energy
Research of the U.\ S.\ Department of Energy.
%
%-------------------------------------------------------------
%

%

\begin{thebibliography}{99}

\bibitem{Blanchard}
{\it Decoherence: Theoretical, Experimental and Conceptual Problems\/},
eds.\ Ph. Blanchard, D. Giulini, E. Joos, C. Kiefer, I.-O. Stamatescu,
(Springer-Verlag, Berlin, Heidelberg, New York, 2000).
%\bibitem{Giulini}
%D. Giulini, E. Joos, C. Kiefer, J. Kupsch, I.-O. Stamatescu, H. D. Zeh,
%{\it Decoherence and the Appearance of a Classical World in Quantum Theory \/}
%(Springer-Verlag, Berlin, Heidelberg, New York, 1996).
\bibitem{Zurek}
W. H. Zurek, Phys. Rev. D {\bf 24}, 1516 (1981);  Phys. Rev. D {\bf 26}, 1862 (1982).
\bibitem{Joos}
E. Joos and H. D. Zeh,
Z. Phys. B {\bf 59}, 223 (1985).
\bibitem{Legget}
A. J. Leggett, S. Chakravarty, A. T. Dorsey, M. P. A. Fisher, A. Garg, and W. Zwerger,
Rev. Mod. Phys. {\bf 59}, 1 (1987).
\bibitem{qtm94} {\it Quantum Tunneling of Magnetization --- QTM'94}\/, eds. L. Gunther and B. Barbara,
  NATO ASI Ser. E, Vol. 301 (Kluwer, Dordrecht, 1995).
\bibitem{kondo}
M. I. Katsnelson, V. V. Dobrovitski, H. A. De Raedt, and B. N. Harmon,
%  ``Destruction of the Kondo effect by a local measurement'',
cond-mat/0205540.

\bibitem{Nielsen}
M. A. Nielsen, I. L. Chuang,
{\it Quantum computation and quantum information\/}
(Cambridge University Press, Cambridge, New York, 2000).

%\bibitem{2} D. P. DiVincenzo, ``The physical implementation of quantum
%  computation'', quant-ph/0002077.
%\bibitem{6} N. Gershenfeld and I. Chuang, Science {\bf 275}, 350 (1997); 
%  I. L. Chuang, N. Gershenfeld, and M. Kubinec, Phys. Rev. Lett. 
%  {\bf 80}, 3408 (1998).
%\bibitem{7} U. M. K. Vandersypen, M. Steffen, G. Breyta, C. S. Yannoni,
%  M. H. Sherwood, and I. L. Chuang, Nature {\bf 414}, 883 (2001).
%\bibitem{8} J. Preskill, Proc. R. Soc. London, Ser. A {\bf 454}, 385 (1998).

%
% Suzuki
%\bibitem{Trotter59} H.F. Trotter,
%Proc. Am. Math. Soc. {\bf 10}, 545 (1959).

\bibitem{Suzuki77}
M. Suzuki, S. Miyashita, and A. Kuroda,
Prog. Theor. Phys. {\bf 58}, 1377 (1977).

\bibitem{DeRaedt87} H. De Raedt, Comp. Phys. Rep. {\bf 7}, 1 (1987).

\bibitem{Suzuki8591} M. Suzuki, J. Math. Phys. {\bf 26}, 601 (1985);
{\it ibid} {\bf 32} 400 (1991).

\bibitem{hdr}
P. de Vries and H. De Raedt,
Phys. Rev. B{\bf 47}, 7929 (1993).

\bibitem{Krech98}
M. Krech, A. Bunker, and D.P. Landau,
Comp. Phys. Comm. {\bf 111}, 1 (1998).

%\bibitem{Kobayashi94} H. Kobayashi, N. Hatano, and M. Suzuki,
%Physica A {\bf 211}, 234 (1994).

\bibitem{qce}
H. De Raedt, A.H. Hams, K. Michielsen, and K. De Raedt,
Comp. Phys. Comm. {\bf 132}, 1 (2000).

%
%\bibitem{Rouhi95} A. Rouhi, J. Wright,
%Comp. in Phys. {\bf 9}, 554 (1995).
%
%\bibitem{Shadwick97} B.A. Shadwick and W.F. Buell,
%Phys. Rev. Lett. {\bf 79}, 5189 (1997).
%
%
%\bibitem{Tran98} P. Tran, Phys. Rev. E {\bf 58}, 8049 (1998).
%
%\bibitem{Michielsen98}
%K. Michielsen, H. De Raedt, J. Przeslawski, and N. Garcia,
%Phys. Rep. {\bf 304}, 89 (1998).
%
%
%\bibitem{DeRaedt83}
%H. De Raedt and B. De Raedt, Phys. Rev. A {\bf28}, 3575 (1983).
%
%
%\bibitem{DeRaedt94}
%H. De Raedt, K. Michielsen, Comp. in Phys. {\bf 8}, 600 (1994).
%
%Cheby
\bibitem{TAL-EZER}
H. Tal-Ezer and R. Kosloff,
J. Chem. Phys. {\bf 81}, 3967 (1984).
%
%\bibitem{SILVER}
%R.N. Silver and H. R\"oder,
%Phys. Rev. E {\bf 56}, 4822 (1997).


%\bibitem{LOH}
%Y.L. Loh, S.N. Taraskin, and S.R. Elliot,
%Phys. Rev. Lett. {\bf 84}, 2290 (2000).
%
%SIL+Cheby
\bibitem{LEFOR}
C. Leforestier, R.H. Bisseling, C. Cerjan, M.D. Feit, R. Friesner, A. Guldberg,
A. Hammerich, G. Jolicard, W. Karrlein, H.-D. Meyer, N. Lipkin,
O. Roncero, and R. Kosloff,
J. Comp. Phys. {\bf 94}, 59 (1991).

\bibitem{Iitaka01}
T. Iitaka, S. Nomura, H. Hirayama, X. Zhao, Y. Aoyagi, and T. Sugano,
Phys. Rev. E {\bf 56}, 1222 (1997).

\bibitem{Slava}
V.V. Dobrovitski and H.A. De Raedt,
preprint submitted to PRE; arXiv: quant-ph/0301130.

%Lanczos
\bibitem{Park}
T.J. Park and J.C. Light,
J. Chem. Phys. {\bf 85}, 5870 (1986).
\bibitem{Manthe}
U. Manthe, H. K\"oppel, and L.S. Cederbaum,
J. Chem. Phys. {\bf 95}, 1708 (1991).
\bibitem{Jaklic94}
J. Jackli\v{c} and P.Prelov\v{s}ek,
Phys. Rev. B. {\bf 49}, 5065 (1994);
%\bibitem{Jaklic00}
%J. Jackli\v{c} and P.Prelov\v{s}ek,
Adv. Phys. {\bf 49}, 1 (2000).
%

\bibitem{Akakii}
A. Melikidze, V.V. Dobrovitski, H.A. De Raedt, M.I. Katsnelson, and B.N. Harmon,
arXiv: quant-ph/0212097.
\bibitem{Slava01}
V. V. Dobrovitski, H. A. De Raedt, M. I. Katsnelson, and B. N. Harmon,
arXiv: quant-ph/0112053.

\bibitem{Wilkinson}
J.H.~Wilkinson, {\sl The Algebraic Eigenvalue Problem}, (Clarendon Press, Oxford, 1965).

%\bibitem{ABRAMOWITZ}
%M. Abramowitz and I. Stegun, Handbook of Mathematical Functions, (Dover,
%New York, 1964).%


%\bibitem{Smith85} G.D. Smith,
%{\it Numerical solution of partial differential equations},
%(Clarendon, Oxford, 1985).
%
\end{thebibliography}
\end{document}